\documentclass[showpacs,prl,twocolumn,superscriptaddress]{revtex4}
\usepackage{graphicx}
\newcommand{\etal}{{\it et al.}}

\begin{document}

\title{Vector Optical Activity in the Weyl Semimetal TaAs}

\author{M. R. Norman}
\affiliation{Materials Science Division, Argonne National Laboratory, Argonne, IL 60439, USA}

\begin{abstract}
It is shown that the Weyl semimetal TaAs can have a significant polar vector contribution to its
optical activity.  This is quantified by ab initio calculations of the resonant x-ray diffraction
at the Ta L1 edge.  For the Bragg vector (400), this polar vector contribution to the circular
intensity differential between left and right polarized x-rays is predicted to be comparable to that 
arising from linear dichroism.  Implications this result has in regards to optical effects 
predicted for topological Weyl semimetals are discussed.
\end{abstract}

\date{\today}

\maketitle

Weyl semimetals are predicted to have a variety of novel optical effects due to their topological
electronic structure \cite{hosur,goswami,kargarian,zhong}.
But in the case of Weyl semimetals which exist because of inversion
symmetry breaking, unusual optical effects can also arise depending on the space group of the lattice.
Disentangling these two sources will be important in order to ascertain which effects arise
due to such phenomena as the axial anomaly, and which are simply due to the influence of the crystallography
on the electronic structure.

Recently, much attention has focused on the Weyl semimetal TaAs where unusual Fermi arc surface
states have been observed by photoemission \cite{xu,lv} as predicted by theory \cite{huang,weng}.
These and related materials also show novel magnetoresistance phenomena, including evidence
for the axial anomaly \cite{zhang,huang2} predicted long ago by Nielsen and Ninomiya \cite{nielsen}.
This axial anomaly can also cause circular dichroism and related chiral optical
effects \cite{hosur,goswami,kargarian,zhong}.

Of course, multiferroics can also exhibit similar optical effects, but perhaps more relevant for the case
of TaAs, chiral crystal structures can as well.  For the latter, these are reciprocal (natural) optical
activity, as opposed to non-reciprocal activity due to time reversal symmetry breaking.  This requires
the breaking of inversion symmetry.  Depending on the space group, a variety of effects can be
observed, and this was spelled out in a classic paper by Jerphagnon and Chemla \cite{jerphagnon}.
The gyration tensor has nine elements that can be decomposed in terms of a pseudoscalar,
a polar vector, and a symmetric traceless second rank tensor known as a pseudodeviator.  The pseudoscalar
is responsible for natural circular dichroism in the optical frequency range due to interference between
the electric dipole and magnetic dipole scattering terms (E1-M1).  The pseudodeviator is responsible for natural 
circular dichroism in the x-ray regime (XNCD) due to interference between the electric dipole and electric 
quadrupole scattering terms (E1-E2) \cite{alagna}.

The second, polar vector contribution, does not lead to optical activity in the traditional sense, but it does lead to
the generation of a longitudinal electric field in the sample \cite{jerphagnon}.  This field, though, is predicted to be
small and thus difficult to observe.  But long ago, Voigt \cite{voigt} and Fedorov \cite{fedorov}
realized this this could lead to polarization rotation
in the reflected light for incoming light not along the normal to the surface.
In 1978, this effect was reported for CdS \cite{ivchenko1,ivchenko2}.
A general theory for this and related optical effects was worked out by Graham and Raab \cite{graham}.
A related polarization rotation has been observed in the x-ray regime at the Zn K edge for ZnO \cite{goulon}.

An important point about CdS, ZnO, and related materials is that only the polar vector contribution is
present.  Interestingly, the $I4_1md$ space group for TaAs and its relatives (TaP, NbAs, NbP) also has this
property \cite{norman}.  This is of particular relevance since similar optical effects have been discussed by
Kargarian \etal~\cite{kargarian} that are connected with the Fermi surface arc states which are known to be
present in TaAs.  In one of their geometries where the surface contains Fermi arcs, they predict the
generation of a longitudinal electric field.  As commented by this author \cite{norman}, for oblique incidence,
one would also expect polarization rotation in the reflected light as in CdS.  Differentiating their topological effect
from the crystallographic effect could be a challenge in materials 
like TaAs.

To get a handle on the latter, we turn to ab initio work.  Although the calculation of optical spectra is very
sensitive to the assumed band structure, this simplifies considerably in the x-ray regime.  As in the work
done on ZnO \cite{goulon}, the approach is to find the optimal conditions to detect the polar vector
contribution.  To see this, we first outline the geometry of such experiments in Fig.~1.  Here, the
surface normal defines the scattering vector, $Q$, which is the difference of the outgoing wavevector $k_o$
and the incoming wavevector $k_i$.
$\theta$ is the Bragg angle which is the angle of $k_i$ relative to the surface (so
an angle of 90$^\circ$ corresponds to normal incidence).  $\psi$ is the azimuthal angle for rotation about $Q$.
$\psi=0^\circ$ corresponds to the incidence plane being defined by $Q$ and $I$, with $I \equiv z$ for $Q$ along $x$,
where $z$ is the optical axis (the $c$ axis in the case of TaAs).  Note that the electric polarization vector for $\sigma$
polarization is perpendicular to the incidence plane, whereas for $\pi$ polarization it is in this plane.

\begin{figure}
\includegraphics[width=0.8\hsize]{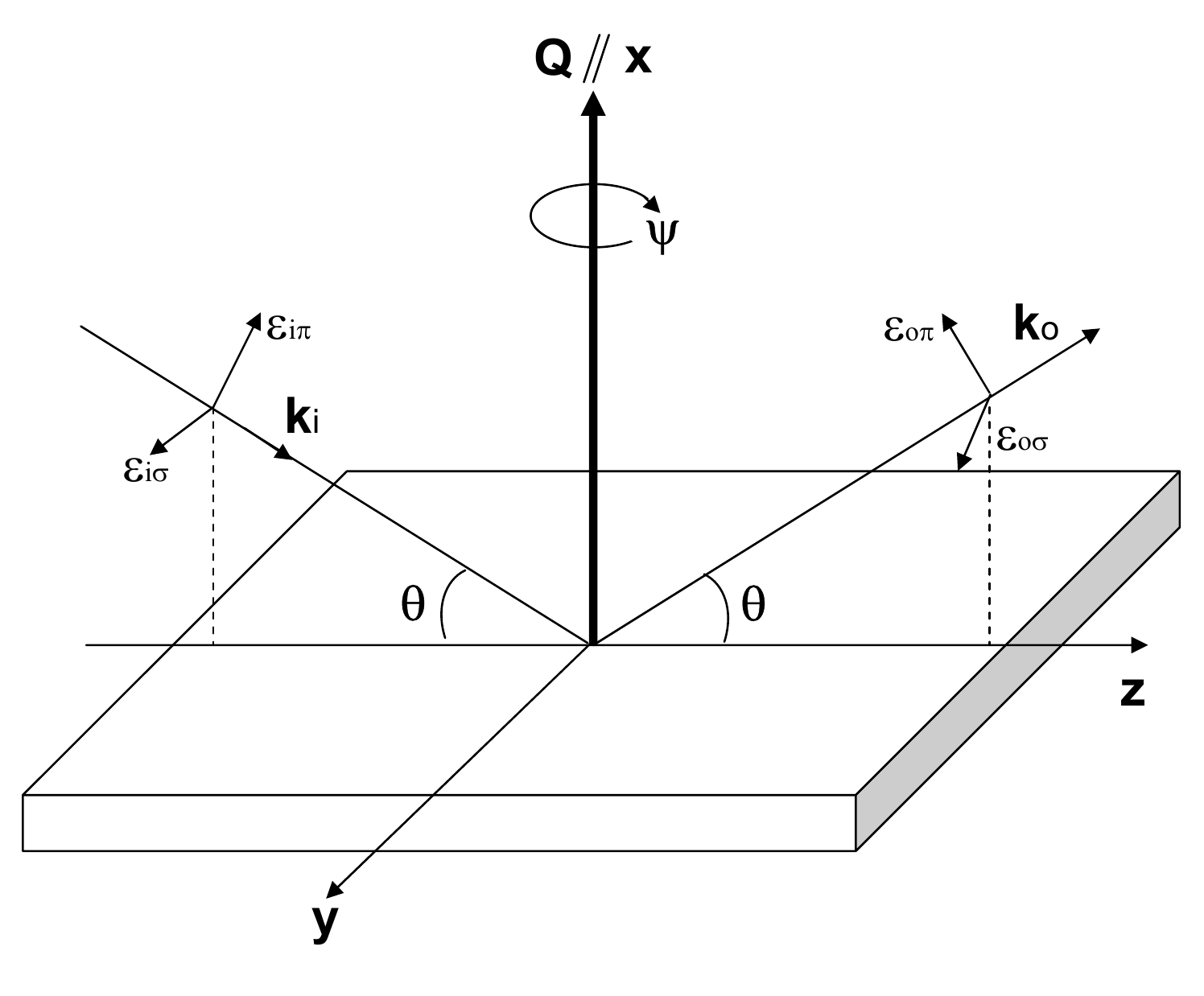}
\caption{Scattering geometry, with $k_i$ the incident wavevector, $k_o$ the outgoing one, and the scattering
vector $Q$ is normal to the surface.  $\theta$ is the Bragg angle and $\psi$ the azimuthal angle, and $\epsilon$
denotes the polarization.  The geometry shown is for $Q$ along $x$, with the optical ($c$) axis along $z$.}
\label{fig1}
\end{figure}

To get at the polar vector optical activity contribution, one can look at the so-called x-ray circular intensity
differential (XCID) which is the difference of scattering intensity between left polarized incoming
x-rays ($L$) and right polarized incoming x-rays ($R$).  With no analysis of the outgoing polarization, this can
be written as \cite{goulon}:
\begin{equation}
I^L - I^R = -2 \rm{Im}[f^{\sigma\sigma}(f^{\pi\sigma})^* + f^{\sigma\pi}(f^{\pi\pi})^*]
\end{equation}
where $f$ is the elastic scattering factor, with the first index denoting incoming polarization, and the last
outgoing polarization.  Note that the total scattering factor is
\begin{equation}
F = \sum_n e^{i Q \cdot r_n} f_n
\end{equation}
where $n$, the site index, has been suppressed in the first equation, noting that the intensity is the 
modulus squared of $F$.
The scattering factor $f$ is a sum of the Thomson scattering, and then various terms corresponding for
resonant scattering to excitations from a core orbital to unoccupied valence orbitals and then back 
again \cite{JMB}.  That is, the scattering matrix elements, $<M|\hat{O}^*|N><N|\hat{O}|M>$
(where $M$ is the ground state and $N$ the excited state) can be expanded in a multipole series
since for the relevant Hamiltonian ($\hat{H}$) terms, $\hat{O} \equiv e^{ik \cdot r} \epsilon \cdot r$ where $k$
is the wavevector and $\epsilon$ the polarization.  This leads to the dipole (E1) contribution $\epsilon \cdot r$
and the quadrupole (E2) contribution $(k \cdot r) (\epsilon \cdot r)$, giving rise in $f$ to
dipole terms (E1-E1), quadrupole terms (E2-E2), and dipole-quadrupole interference terms (E1-E2), the last
existing only if the site $n$ does not have inversion symmetry.  Here, additional magnetic dipole terms coming from
$\hat{H}$ have been dropped since they are negligible in the x-ray regime, as well as higher order (octupole) terms
from the expansion of $\hat{O}$.

As discussed by Graham and Raab \cite{graham}, the desired polar vector effect cannot be observed by XCID
with $Q$ along the optic axis, though a related intensity differential can occur (see below).
Instead, we first turn to the case when $Q$ is along the $x$ direction as in the work on
ZnO \cite{goulon}, where by $x$ we mean along the tetragonal $a$ axis.

For the polar vector contribution to the optical activity, the crucial term is the E1-E2 contribution to the
scattering factors $f^{\sigma\pi}$ and $f^{\pi\sigma}$, which was shown
by Goulon \etal~\cite{goulon} for the point group 6mm and $Q$ along $x$ to be proportional to
$\sin(2\theta) \sin(\psi) t_{xxz}$ where $t_{xxz} \equiv <M|r_x|N><N|r_xr_z|M>$, which can be easily derived
from the functional form for E1-E2 \cite{goulon,JMB} of
$\sum_{\alpha\beta\gamma} \epsilon_{o\alpha}^* \epsilon_{i\beta}
(t_{\alpha\beta\gamma}k_{i\gamma} - t_{\beta\alpha\gamma} k_{o\gamma})$.
This contribution (zero for $\sigma\sigma$ and $\pi\pi$) is invariant under interchange of $\sigma$ and $\pi$ indices.
On the other hand, there are dipole (E1-E1) and quadrupole
(E2-E2) contributions to these two scattering factors as well (noting that $f^{\sigma\sigma}$ and $f^{\pi\pi}$
are dominated by the large Thomson scattering term which does not contribute to
$f^{\sigma\pi}$ and $f^{\pi\sigma}$).  The dipole one
($\sum_{\alpha\beta} \epsilon_{o\alpha}^* \epsilon_{i\beta} d_{\alpha\beta}$) goes as
$\sin(\theta) \sin(2\psi) (d_{zz}-d_{xx})$ where $d_{ii} \equiv <M|r_i|N><N|r_i|M>$,
with this contribution being odd under the interchange of $\sigma$ and $\pi$ indices.
The more complicated quadrupole (E2-E2) term instead involves the azimuthal factor $\sin(4\psi)$.
These forms can be easily shown to apply to the 4mm point group of TaAs as well.  
Because of the differing azimuthal factors of these three terms, they can be
differentiated by performing an azimuthal sweep.  In particular, for an azimuthal angle of 90$^\circ$ (that is,
with the $c$ axis perpendicular to the incidence plane), the E1-E1 and E2-E2 terms vanish, and the XCID
is determined by the polar vector E1-E2 contribution.

Based on the above, to maximize this polar vector contribution, one
wants Bragg angles near 45$^\circ$ \cite{goulon}.
At the Ta L1 edge, the Bragg vector (400) (2$\pi/a$ units) has a Bragg angle of 38.1$^\circ$,
close to the desired value.  To proceed, we turn to ab initio work, employing the multiple scattering Greens
function code FDMNES \cite{joly} including spin-orbit interactions \cite{so}.
The simulations were done using local density (LDA) atomic potentials (Hedin-Lundqvist 
exchange-correlation function) in a muffin tin approximation that considers multiple scattering of the
photoelectron around the absorbing site \cite{natoli}.  The cluster radius is limited
by the photoelectron lifetime \cite{escape}. 
For the present case (Ta L1 edge), the results for cluster radii of 5 \AA~and 6 \AA~are similar,
indicating cluster convergence.
Results shown are for a radius of 6 \AA, which corresponds to 55 atoms around the Ta site.

Fig.~2a shows the x-ray absorption spectrum for TaAs at the L1 edge.  From this plot, one can see that the
x-ray linear dichroism for this material is predicted to be weak.  The resulting resonant x-ray scattering
intensity for the Bragg vector (400) is shown in Fig.~2b for incoming right and left polarized light (summed over
outgoing polarizations).  Again, these two spectra are almost identical since the scattering is dominated by
the Thomson scattering term which is large for Ta, which has a large $Z$.  Subtracting the two polarizations,
one obtains the x-ray circular intensity differential (XCID) shown in Fig.~2c for several representative
azimuthal angles.  The spectra for 30$^\circ$ and 60$^\circ$ are similar, but the one at 90$^\circ$ is different.
For a representative energy, the azimuthal dependence of the intensity is plotted in Fig.~2d.  This can be
fit by the sum of three terms, one going as $\sin(\psi$), the others as $\sin(2\psi)$ and $\sin(4\psi)$.  The latter
two are due to the E1-E1 and E2-E2 terms and are related to the x-ray linear dichroism (XLD), as shown more
explicitly in Fig.~3a for an azimuthal angle of 45$^\circ$.  The first, though, is due to the desired x-ray optical
activity (XOA) coming from the E1-E2 interference term $t_{xxz}$.  This term determines the XCID
at an azimuthal angle of 90$^\circ$ as seen in Fig.~2d and illustrated further in Fig.~3b.  For a general azimuthal 
angle, the XOA contribution is predicted to be significant compared to the  XLD
contribution, as contrasted with ZnO at the Zn K edge \cite{goulon}.  On the other hand, in absolute numbers,
the XOA term is small, of order 0.03\% of the total scattering intensity shown in Fig.~2b.

\begin{figure}
\includegraphics[width=\hsize]{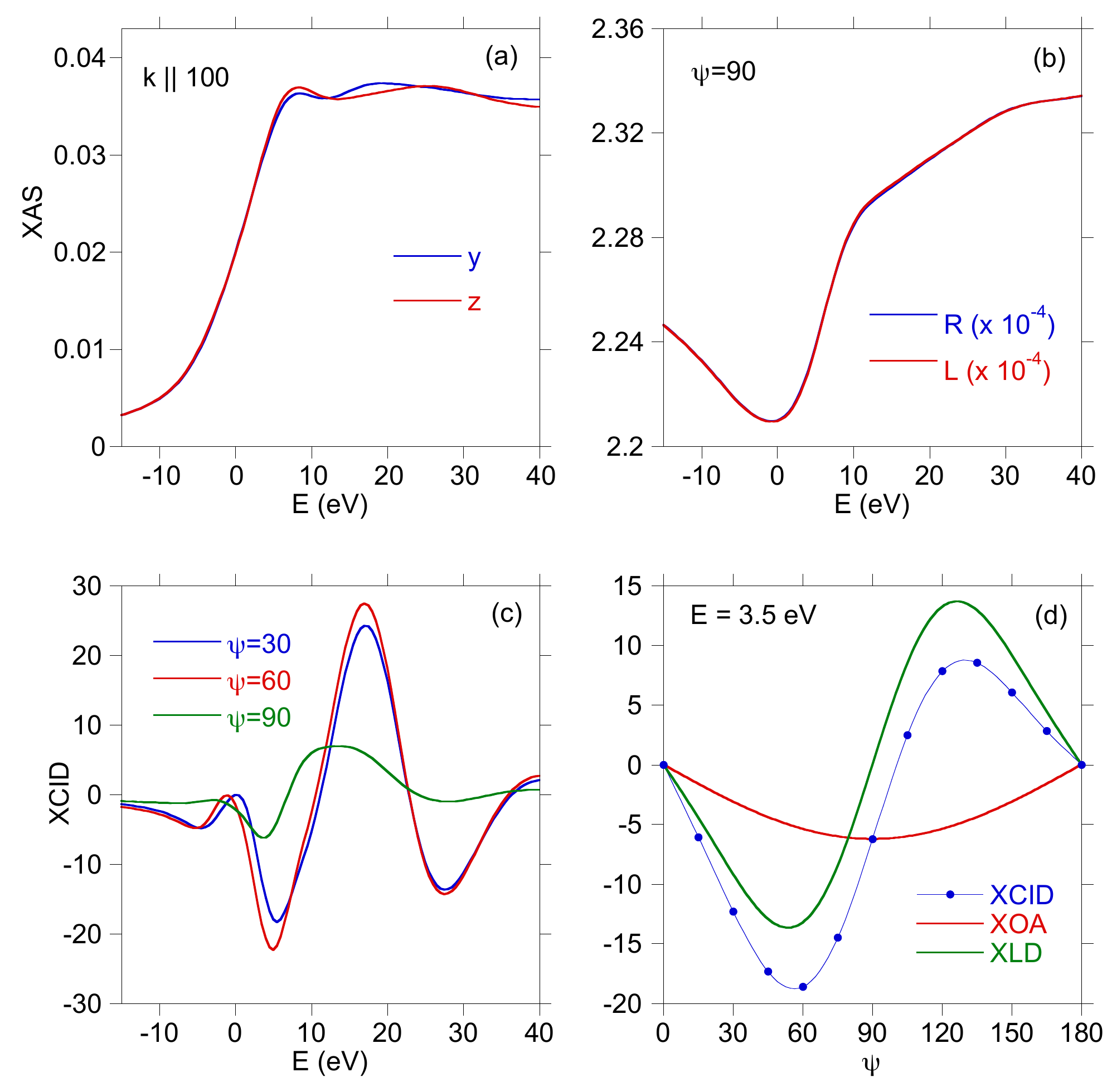}
\caption{(Color online) Ta L1 edge - (a) X-ray absorption (XAS) for $k$ along (100) for electric polarizations
parallel to $y$ and $z$. (b) Resonant x-ray scattering intensity for incoming left (L) and right (R) polarized x-rays
for a Bragg vector (400) and an azimuthal angle of 90$^\circ$. (c) XCID intensity (L - R) for three azimuthal
angles. (d) XCID intensity and corresponding decomposition into XLD and XOA contributions at an energy of
3.5 eV.  The unit for absorption is Mbarn, and for scattering intensities number of electrons squared (summed
over the unit cell).  The zero of energy is at 11.682 keV.}
\label{fig2}
\end{figure}

This brings us to the question of whether there is something else besides the XCID that could be exploited.
The answer is yes if one has control over both incoming polarization and measuring outgoing polarization.
The reason is that the Thomson scattering does not contribute to the $\sigma\pi$ and $\pi\sigma$ terms.  In
Fig.~3c, the scattering intensity for these polarization settings are shown for two representative azimuthal angles.
Again, at 90$^\circ$, the total contribution is due to the XOA one, with the large difference in the two polarization
settings for 45$^\circ$ again due to the XOA term (Figs.~3c and 3d).

\begin{figure}
\includegraphics[width=\hsize]{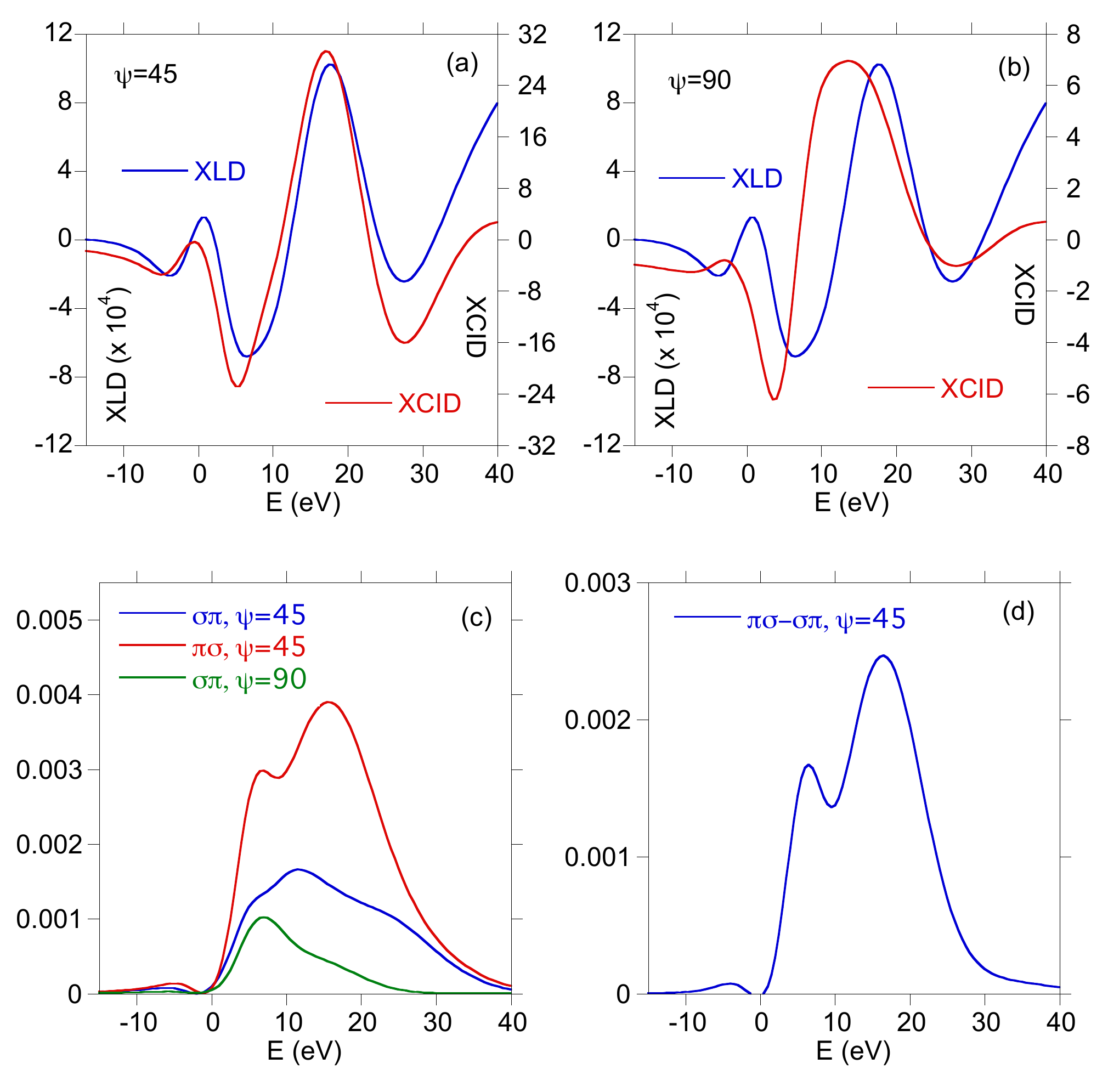}
\caption{(Color online) Ta L1 edge - Comparison of XLD ($y-z$) and XCID for an azimuthal angle of 
(a) 45$^\circ$
and (b) 90$^\circ$. (c) Resonant x-ray intensities for various azimuthal angles, with the first polarization index for
incoming x-rays, the second for outgoing x-rays. (d) Difference of the two intensities in (c) for an azimuthal
angle of 45$^\circ$ (for 90$^\circ$, the difference is zero).}
\label{fig3}
\end{figure}

So, what about Bragg vectors along the $c$ axis?  Calculations have also been done at the Ta M1 edge
for $Q$=(004) (2$\pi/c$ units).  As discussed by Graham and Raab \cite{graham}, one way to get at the polar vector term in this 
case is to look at incoming light (either $L$ or $R$), but measuring the outgoing light for 45$^\circ$ 
and -45$^\circ$ polarizations and taking the difference (with $\pi$ corresponding to 0$^\circ$ and $\sigma$ to 
90$^\circ$).  This is challenging, since as in the previous paragraph, it requires exquisite control of the incoming 
polarization and measuring the outgoing one.  But if one calculates the azimuthal dependence, one finds that
both the E1-E1 and E1-E2 contributions do not depend on azimuthal angle in this geometry, and therefore it is
the quadrupole terms that solely drives this dependence (which is predicted to be weak).  Therefore, although
XOA does exist for this geometry, it would be difficult to quantify experimentally, which can be traced to the fact 
that in this geometry, the $\sigma\pi$ and $\pi\sigma$ terms vanish identically \cite{graham}.

We now turn to the implications our results have in regards to optical activity due to the topological properties
of this material, which played no role in the above calculations.  This has been treated most definitively by
Kargarian \etal~\cite{kargarian}.  To make connection to this work, let us summarize what would be expected 
for optical activity due to crystallographic effects.  For the point group relevant for TaAs, a longitudinal electric
field inside the sample is expected for light propagating along the $x$ ($a$) axis and polarization along the $z$ ($c$) axis 
due to the polar vector optical activity \cite{jerphagnon}.  This can be easily seen since in this case, the only
non-zero terms of the gyration tensor are $g_{yx}=-g_{xy}$.  This is analogous to the longitudinal electric
field discussed by Kagarian \etal~\cite{kargarian} in a geometry where the surface contains Fermi arcs.
The associated optical activity can be determined by reflection if the incidence wavevector is not along the
surface normal.  According to Graham and Raab \cite{graham}, this shows up as a polarization rotation if the
optic axis ($c$ axis in the present case) is in the surface and also not in the incidence plane (that is $\psi$
not equal to zero in the geometry of Fig.~1).  The analogous XOA results are shown in Figs.~2 and 3.
These XOA effects also contribute to $\sigma\sigma$ and $\pi\pi$ if the optic axis instead is along the surface normal,
but as discussed in the previous section, they are not easily separated from the ordinary dipole contribution
since neither depends on the azimuthal angle.  Interestingly this is exactly the surface for TaAs that contains Fermi 
arcs \cite{xu,lv,huang,weng}.  For the geometry appropriate to Figs.~2 and 3 (that is, with the surface normal along
the $a$ axis), the situation is less clear, since
such a surface may or may not contain Fermi arcs \cite{weng}, and the evidence from photoemission either way is not
clear \cite{xu}.  Certainly, we anticipate that what is due to crystallography, and what is due to the topological electronic
structure, may be difficult to separate.

On the other hand, if the topological effect is due to time reversal breaking \cite{kargarian}, then the two effects 
can in principle be distinguished by applying a magnetic field.  An analogous effect has been demonstrated in 
tellurium, which is also thought to be topological in nature \cite{hirayama}.  At zero field, polarization rotation
occurs for transmitted light due to fact that the space group breaks inversion symmetry (with both natural optical 
activity and XNCD allowed \cite{norman}), but the application of a current leads to an additional polarization
rotation due to time reversal breaking \cite{vorobev}.  Such a Faraday rotation is predicted by 
Kargarian \etal~\cite{kargarian} for a surface which does not contain Fermi arcs, but interestingly no optical
rotation would occur in TaAs due to crystallography since its space group does not allow for natural optical
activity (unlike the case for tellurium) \cite{norman}.  So, in this case, any polarization rotation of the transmitted
light should be topological in nature.

In summary, Weyl semimetals can exhibit optical activity due both to its topological electronic structure,
and to crystallography.  By constructing experiments where the latter effect is minimized, the unique
topological signatures can be identified.  Regardless, the novel Weyl semimetal TaAs, as well as its related
siblings (TaP, NbAs, NbP), should exhibit novel optical activity of a polar vector nature that in principle
can be identified by appropriate resonant x-ray diffraction measurements, as demonstrated here.

This work was supported by the Materials Sciences and Engineering
Division, Basic Energy Sciences, Office of Science, US DOE.


\begin{thebibliography}{99}
\bibitem{escape}
The results presented involve a convolution of the calculated spectrum with 
both a core hole (7.68 eV for the Ta L1 edge) and a photoelectron
inverse lifetime, with the latter having a strong energy dependence (at high energies, 15 eV, with a
midpoint value at 30 eV above the Fermi energy) \cite{so}.  Core hole widths tabulated
in the FDMNES program are based on the x-ray literature \cite{so}.
\end{thebibliography}
\end{document}